\documentclass[twocolumn,prb,amsmath,amssymb,floatfix,superscriptaddress,showpacs]{revtex4}
\usepackage{graphicx}
\usepackage{dcolumn}
\usepackage{bm}
\usepackage{times}
\begin{document}


\title{Disappearance of static magnetic order and evolution of spin fluctuations
in Fe$_{1+\delta}$Se$_{x}$Te$_{1-x}$ }

\author{Zhijun Xu}
\affiliation{Condensed Matter Physics \&\ Materials Science
Department, Brookhaven National Laboratory, Upton, New York 11973,
USA} \affiliation{Department of Physics, City College of New York,
New York, New York 10033, USA}
\author{Jinsheng Wen}
\affiliation{Condensed Matter Physics \&\ Materials Science
Department, Brookhaven National Laboratory, Upton, New York 11973,
USA}\affiliation{Department of Materials Science and Engineering,
Stony Brook University, Stony Brook, New York 11794, USA}
\author{Guangyong Xu}
\affiliation{Condensed Matter Physics \&\ Materials Science
Department, Brookhaven National Laboratory, Upton, New York 11973,
USA}
\author{Qing Jie}
\affiliation{Condensed Matter Physics \&\ Materials Science
Department, Brookhaven National Laboratory, Upton, New York 11973,
USA}\affiliation{Department of Materials Science and Engineering,
Stony Brook University, Stony Brook, New York 11794, USA}
\author{Zhiwei Lin}
\author{Qiang Li}
\affiliation{Condensed Matter Physics \&\ Materials Science
Department, Brookhaven National Laboratory, Upton, New York 11973,
USA}
\author{Songxue Chi}
\author{D. K. Singh}
\affiliation{NIST Center for Neutron Research, National Institute of
Standards and Technology, Gaithersburg, Maryland 20899,
USA}\affiliation{Department of Materials science and Engineering,
University of Maryland, College Park, Maryland 20742, USA}
\author{Genda Gu}
\author{J. M. Tranquada}
\affiliation{Condensed Matter Physics \&\ Materials Science
Department, Brookhaven National Laboratory, Upton, New York 11973,
USA}
\date{\today}

\begin{abstract}
We report neutron scattering studies on static magnetic orders and
spin excitations in the Fe-based chalcogenide system
Fe$_{1+\delta}$Se$_{x}$Te$_{1-x}$ with different Fe and Se compositions.
Short-range static magnetic order with the ``bicollinear'' spin configuration
is found in all non-superconducting samples, with strong low-energy
magnetic excitations near the $(0.5,0)$ in-plane wave-vector (using the
two-Fe unit cell) for Se doping up to 45\%. When the static order disappears and bulk superconductivity emerges, the spectral weight of the magnetic excitations shifts to the region of reciprocal space near the in-plane wave-vector $(0.5,0.5)$, corresponding to the ``collinear'' spin configuration. Our results suggest that spin
fluctuations associated with the collinear magnetic structure appear to be
universal in all Fe-based superconductors, and there is a strong
correlation between superconductivity and the character of the magnetic order/fluctuations in
this system.

\end{abstract}


\maketitle

\section{Introduction}

Since the discovery of the first high temperature
superconductor in the 1980's,  there has been a continuing effort to understand the origin of high-$T_c$
superconductivity. Studies on the cuprate systems seem to suggest
that there is an intimate relationship between superconductivity and
magnetism,~\cite{Tranquada2004,Fong1999,Xug2009np,Mook1998} and recently this
has been shown to be the case also for the newly discovered
Fe-based superconductor
families.~\cite{Kamihara2008,Takahashi2008,Chen2008,Ren2008,Hsu2008,Yeh2008}
With non-superconducting parent compounds that have static antiferromagnetic (AF) order, charge doping in the Fe pnictides
gradually suppress the AF order and induces superconductivity, for
both the RFeAsO (``1111''
system)~\cite{Clarina2008,Zhaoj2008nm,Huang2008prb,Lee2008jpsj,Zhaoj2008prb,Qiu2008prb}
and AFe$_2$As$_2$ (``122'' system)
families.~\cite{Zhaoj2009nf,Chen2009epl} In addition, a ``spin
resonance'' has been observed in the ``122'' system by inelastic
neutron
scattering,~\cite{Christianson2008,Lumsden2009prl,Chis2009prl,Lis2009prb,Inosov2010nf}
showing a sharp increase of  the magnetic scattering intensity at
the ``resonance'' energy when the system goes into the
superconducting phase. For the pnictides, both the static magnetic
order in the parent compound and the ``resonance'' in the
superconducting compounds occur around the in-plane wave
vector (0.5,0.5) (using the 2-Fe unit cell), suggesting a
``collinear'' or ``C-type'' AF
structure [see Fig.~\ref{fig:1} (a)].~\cite{Yinw2010,Clarina2008,Lis2009}

The situation is slightly different in another Fe-based
superconductor family, the iron chalcogenide
Fe$_{1+\delta}$Se$_{x}$Te$_{1-x}$ (the ``1:1'' compound). Here the
parent compound has a ``bicollinear'' or ``E-type'' AF order [see
Fig.~\ref{fig:1} (b)],~\cite{Baow2009prl,Lis2009,Wen2009,Yinw2010}
modulated along the $(0.5,0)$ in-plane direction. Furthermore, the substitution of Se for
Te, which induces superconductivity, does not directly modify the density of electrons in the conduction bands. Despite these
differences, the spin resonance observed in the
superconducting compositions of this family has been found to occur at the same
$(0.5,0.5)$ in-plane
wave-vector,~\cite{Qiu2009,Mook2009,Wen2010H,Argyriou2009,Lee2010,Lis2010,Baow2010}
as in the Fe pnictides, but rotated 45$^\circ$ from the ordering
wave-vector of the parent FeTe compound.  This suggests that the
superconducting mechanism in the Fe pnictides and chalcogenides are
likely to be quite similar. Nevertheless, there has also been a report
implying that superconductivity can coexist with the ``bicollinear''
structure on the atomic scale.~\cite{Khasanov2009} Therefore,
understanding how the magnetic structure (static or dynamic) evolves
from the E-type bicollinear configuration of the parent compound to the
C-type collinear configuration in the superconducting region in the
1:1 system is an important problem. There have been a number of
theoretical studies~\cite{Yinw2010,Fang2009} that provide some insight
on this issue, but clearly a more systematic experimental study is
highly desirable.

In this paper, we present our work using neutron scattering to probe
the magnetic order/fluctuations in a few samples from the 1:1 family for
Se dopings ranging from 30\%\ to 50\%\  and with varying
superconducting properties. Our results suggest that static
magnetic order exists in all non-superconducting samples.  (Here, by non-superconducting we mean an absence of bulk superconductivity.) This order
is short-ranged and occur at in-plane wave-vectors of the type $(0.5,0)$.
For the fully superconducting samples, no static magnetic order is found.
With the disappearance of static magnetic order, the associated low energy magnetic
excitations near $(0.5,0)$ also go away, as one might expect. Magnetic excitations near
$(0.5,0.5)$ gradually become dominant as the material
becomes more superconducting. While Se doping plays an essential role, it is
clearly not the only determining factor regarding the superconductivity and
the magnetic correlations.
Samples with similar Se doping but differing in Fe content can have very different superconducting properties and corresponding magnetic structures/fluctuations. Our results
clearly indicate that static bicollinear (E-type) magnetic orders in the
1:1 system compete with and suppress superconductivity.
Superconductivity only appears when the system evolves toward a
fluctuating collinear C-type magnetic correlations, which appear to
be universal across all known Fe-based superconductor families.

\section{Experiment}

\begin{figure}[ht]
\includegraphics[width=\linewidth]{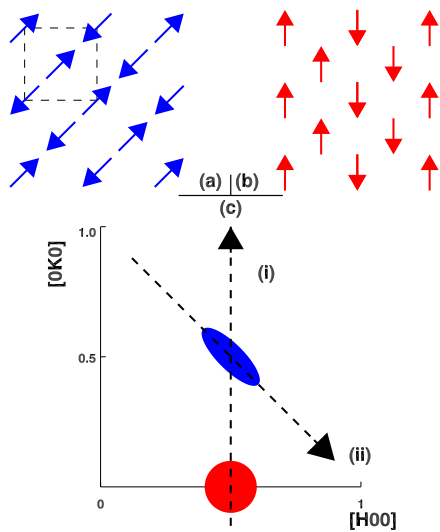}
\caption{(Color online) (a) Schematic of the ``collinear'' C-type AF spin
structure, with scattering intensities mainly around $(0.5,0.5)$.
The square shows a unit cell with two Fe atoms.
(b) Schematic of the ``bicollinear'' E-type AF
spin structure, which contribute mainly to scattering intensities near $(0.5,0)$.
(c) The schematic diagram of the neutron scattering measurements in the $(HK0)$
zone. Dashed lines denote linear scans performed across $(0.5,0.5)$ and
$(0.5,0)$ in the text.}
\label{fig:1}
\end{figure}

\begin{table}
\caption{List of the Fe$_{1+\delta}$Se$_x$Te$_{1-x}$ samples used in our measurements, with their composition ($\delta$, $x$),
superconducting transition temperature ($T_c$), room-temperature lattice parameters (from powder x-ray diffraction), and crystal mass.}
\begin{ruledtabular}
\begin{tabular}{ccccccc}
  Sample & $\delta$ & $x$ & $T_{c}$ & $a$  & $c$ & mass \\
                 &                &         &   (K)       & (\AA) & (\AA) & (g)   \\
  \hline
   SC30 & 0.00 & 0.3 & 14 & 3.815 & 6.140 & 12.7 \\
   NSC30 & 0.05 & 0.3 & --  & 3.808 & 6.120 & 7.4   \\
   SC50 & 0.00 & 0.5 & 14 & 3.811 & 6.129 & 9.0 \\
   NSC45 & 0.05 & 0.45 & --  & 3.807 & 6.047 & 6.4 \\
\end{tabular}
\end{ruledtabular}
\label{tab:1}
\end{table}

\begin{figure}[ht]
\includegraphics[width=\linewidth]{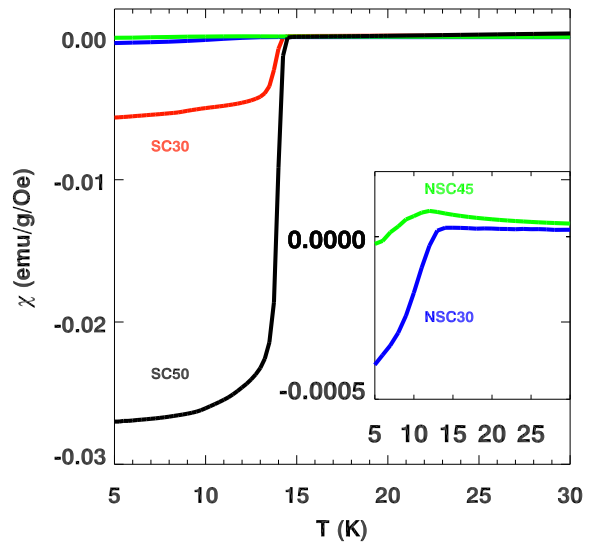}
\caption{(Color online) ZFC magnetization measurements by SQUID for
SC30 (red), NSC30 (blue), SC50 (black) and NSC45 (green). The inset
shows the same data from the non-superconducting samples with
different scale.} \label{fig:2}
\end{figure}

The single-crystal samples used in the experiment were grown by a
unidirectional solidification method. The samples, their nominal compositions, and various characteristic properties are listed in Table I.
The bulk
susceptibility results in Fig.~\ref{fig:2} were obtained using a
superconducting quantum interference device (SQUID) magnetometer.
From the magnetization measurements we can see that, although both
superconducting samples show evidence of diamagnetic response at around $14$~K,
SC50 is
clearly better in quality as far as superconducting volume fraction is concerned. With
a considerable portion of its bulk volume being non-superconducting, it is
possible that there is phase separation in SC30. In fact, when measuring
different small pieces ($\sim 1$~mm size) from the same SC30 sample, the
superconducting volume can vary from $\alt 10$\%\ to $\sim 80$\%, suggesting
that the superconducting and nonsuperconducting phases could be macroscopically
separated in this sample. The other samples, NC30 and
NC45, are mostly non-superconducting, with no more than 1\% of the volume giving a
superconducting response.

Neutron scattering experiments have been carried out on the triple-axis
spectrometers SPINS (inelastic scattering measurements of magnetic excitations
for SC30 and NSC30, and all elastic measurements for static magnetic order),
and BT-7 (inelastic scattering measurements for SC50 and NSC45)
located at the NIST Center for Neutron Research. We used horizontal beam collimations of
Guide-$80'$-S-$80'$-$240'$ (S represents ``sample'') for the inelastic scattering
measurements on SPINS with fixed final energy of 5 meV and a cooled
Be filter after the sample to reduce higher-order neutrons;
collimations of Guide-open-S-$80'$-$240'$ were used for the elastic
measurements on SPINS, with an additional  Be filter  before
the sample. At BT7, we used beam collimations of open-$50'$-S-$50'$-$240'$
with fixed final energy of 14.7 meV and two pyrolytic graphite filters
after the sample.
The inelastic scattering measurements have been performed in the $(HK0)$
scattering plane, as indicated in Fig.~\ref{fig:1}(c). The data are described in reciprocal
lattice units (r.l.u.) of $(a^*, b^*, c^*) = (2\pi/a, 2\pi/b, 2\pi/c)$.
The elastic scattering measurements have been taken in the $(H0L)$ scattering
plane instead, since the order in the parent compound occurs at half
integer $L$ values (AF order along the $c$-axis). All data have been normalized
into absolute units (${\mu_B}^2\cdot$eV$^{-1}/$Fe), using incoherent elastic
scattering intensities from the samples.

\section{Results and Discussion}

\begin{figure}[ht]
\includegraphics[width=\linewidth]{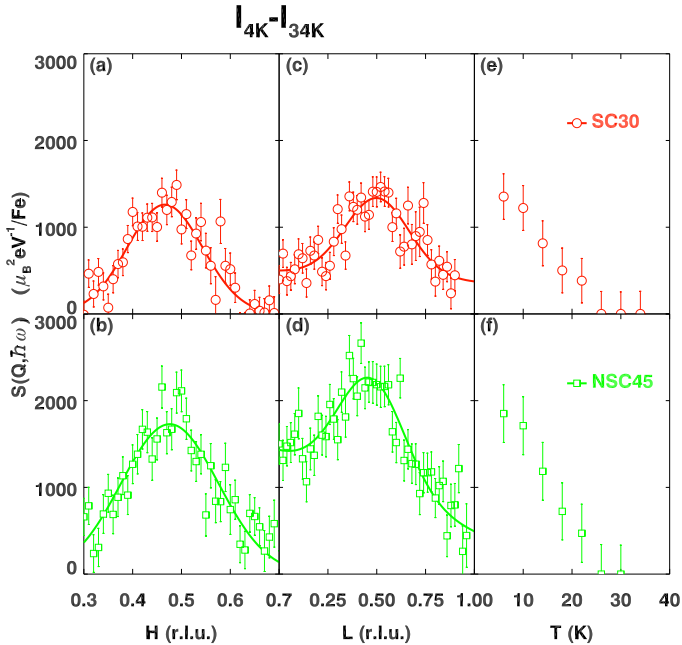}
\caption{(Color online) Elastic neutron scattering measurements
performed on SC30 (top) and NC45 (bottom) near $(0.5,0,0.5)$. (a)
and (b) are intensity profiles along [100] direction ($H$-scans);
(c) and (d) are scans along [001] direction ($L$-scans). (e) and (f)
show the magnetic peak intensity at $(0.5,0,0.5)$ vs. temperature.
Corresponding scans measured at $T=34$~K are used as background, and
have been subtracted from all the data shown. The error bars
represent the square root of the number of counts.} \label{fig:3}
\end{figure}

The static long-range magnetic order in the parent compound FeTe
appears near $(0.5,0,0.5)$, suggesting a bicollinear E-type AF
structure. With small Se doping, it was suggested that the order
should gradually become short-ranged~\cite{Katayama2010,Wen2009} and
eventually disappear. Our results, however, suggest that the order
can still remain with relatively large Se dopings.

We performed elastic magnetic scattering measurements on all
samples. For the SC50 sample, there is no elastic magnetic intensity
at $(0.5,0,0.5)$, while magnetic peaks are observed for all three
other samples. In Fig.~\ref{fig:3}, we plot $H$ and $L$ scans
through this AF wave-vector for the SC30 and NSC45 samples at
$T=4$~K. The same measurements for NSC30 have been published in a
previous report (see Fig.~2 of Ref.~\onlinecite{Wen2009}). The $H$
and $L$ scans performed at higher temperature (T=34~K) show no peak
structure and are therefore used as backgrounds to be subtracted
from the data. All peaks are much broader than the resolution,
indicating the short-ranged nature of the magnetic order. The
$H$-scans are peaked near but not exactly at $H=0.5$, similar to the
results reported for NSC30 and another sample with 27\% Se
doping.~\cite{Wen2009} The $L$-scans, however, are qualitatively
different. In previous reports on lower Se doped 1:1
compounds,~\cite{Lee2010,Wen2009} the $L$-scan peaks around $L=0.5$,
and intensity always goes to zero at $L=0$. So there the magnetic
order is always AF along the $L$-direction, whether short- or
long-ranged. Here we see that after background subtraction, the
scattering intensity at $L=0$ is still appreciable. This suggests
that although the magnetic order still has a modulation along the
$L$-direction, which peaks around $L=0.5$, favoring an AF
configuration between Fe planes, the order has become much more
two-dimensional. In the NSC45 sample, the 3D long-range bicollinear
AF magnetic order of the parent compound has not been entirely
destroyed, but rather greatly reduced to 2D short-range order.  The
ordered moment per Fe is 0.015(4)~$\mu_B^2$, much less than the
value in the parent compound with long-range
order.~\cite{Baow2009prl}  It is nevertheless, enough to destroy
superconductivity. With this static magnetic order present, even
with 45\% Se doping, bulk superconductivity is still not achieved.
In the SC30 sample, although the sample does show a superconducting
phase transition at around 14~K, the superconducting volume is
smaller than for the SC50 sample. The ordered moment is about
0.006(2)~$\mu_B^2/$Fe, also much less than that in the NSC45 sample,
indicating that this order may be coming from only part of the
sample. Therefore there is likely a phase separation in this SC30
sample, where two phases, one superconducting and another one with a
short-range magnetic order, coexist.

\begin{figure}[ht]
\includegraphics[width=\linewidth]{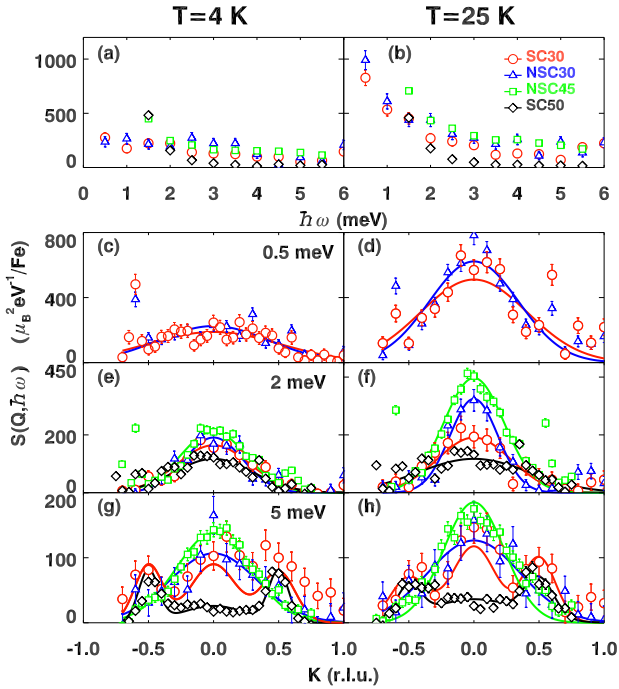}
\caption{(Color online) Magnetic excitations for
Fe$_{1+\delta}$Se$_{x}$Te$_{1-x}$ measured around $(0.5,0,0)$. The
left and right columns show the magnetic peak profiles for lowest
temperature and 25 K respectively. (a) and (b) Constant-$Q$ scans at
$(0.5,0,0)$ taken at low $T$ and 25~K. (c-h) Constant-energy scans
at $(0.5,K,0)$ at $\hbar\omega=$0.5, 2 and 5~meV.  A fitted
$K$-independent background has been subtracted from all data sets.
The error bars represent the square root of the number of counts.}
\label{fig:4}
\end{figure}

With the tendency of forming static bicollinear magnetic structures
in the non-superconducting samples, it is natural to expect to see
magnetic excitation spectra around the (0.5,0) in-plane wave-vector
as well. Previous work has shown that the energy dispersion and
intensity modulation along the $L$ direction for magnetic
excitations in the 1:1 compound is small.~\cite{Lee2010,Qiu2009} We
can therefore choose to perform the inelastic scattering
measurements in the $(HK0)$ plane for $L=0$. In Fig.~\ref{fig:4}, we
plot our results taken near $(0.5,0,0)$. The top two panels
[Fig.~\ref{fig:4} (a) and (b)] show energy scans at $(0.5,0,0)$ at
$T=4$~K and 25~K. Measurements for NSC45 and SC50 were taken on BT7
with a relatively coarse energy resolution (FWHM $\sim$ 1.7meV)
compared to those on SPINS (NSC30 and SC30, FWHM $\sim0.3$~meV), and
have a large, resolution-limited tail from scattering at
$\hbar\omega =0$. Constant-energy scans at $\hbar\omega=0.5$, 2, and
5~meV [Fig.~\ref{fig:4} (c) to (h)] along the $K$ direction across
$(0.5,0,0)$ clearly show that for NSC30, SC30 and NSC45, there is
significant spectral weight at low energies here.
For both 30\%-Se samples, where we have measurements with higher energy resolution, one can see that the intensity at $\hbar\omega=0.5$~meV increases on warming from 4~K to 25~K.
The increase is much less pronounced at
$\hbar\omega=5$~meV.  
This behavior is likely due to a transfer of spectral weight
from the elastic peak into low energy channels when the static order
dissolves with heating. 
For
SC50, the low-energy spin excitations near $(0.5,0,0)$ are weak,
and not strongly temperature dependent, which is consistent with the
fact that there is no static order near (0.5,0) in this sample. The
two small peaks near $K=\pm0.5$ observed from samples SC30 and SC50
at $\hbar\omega=5$~meV, suggest that there is additional spectral weight
developing near the (0.5,0.5) wave-vector, corresponding to dynamic
collinear spin correlations in the superconducting
samples.

\begin{figure}[ht]
\includegraphics[width=\linewidth]{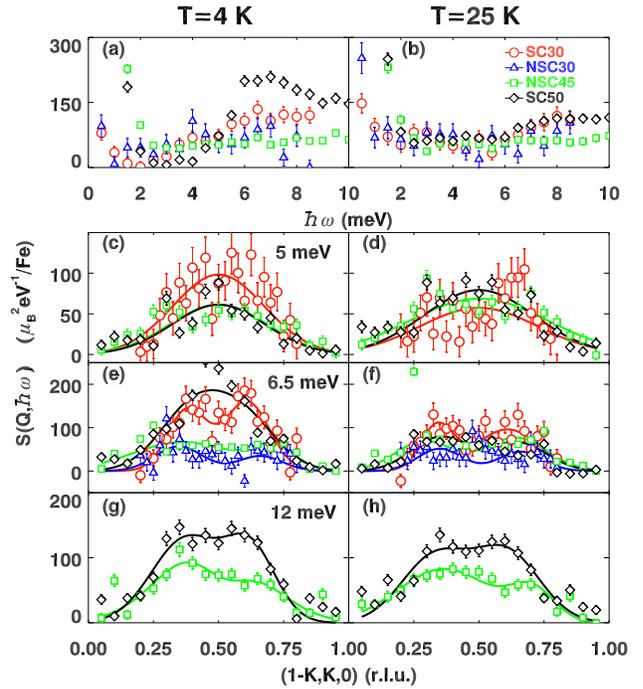}
\caption{(Color online) Magnetic excitations for
Fe$_{1+\delta}$Se$_{x}$Te$_{1-x}$ measured around $(0.5,0.5,0)$. The
left and right columns show the magnetic peak profiles for lowest
temperature and 25 K respectively. (a) and (b) Constant-$Q$ scans at
$(0.5,0.5,0)$ taken at low $T$ and 25~K. (c-h) Constant-energy scans
at $(0.5,0.5,0)$, taken along the transverse direction [as shown in
Fig.~\ref{fig:1} (a)] at $\hbar\omega=5$, 6.5 and 12~meV. A fitted
constant background has been subtracted from all data sets. The
error bars represent the square root of the number of counts.}
\label{fig:5}
\end{figure}

In Fig.~\ref{fig:5}, we show measurements near $(0.5,0.5,0)$. For
SC50, a clear ``resonance'' is observed when comparing the energy
scans performed at 4~K and 25~K [see Fig.~\ref{fig:5} (a) and (b)].
In panels (c) to (h), constant-energy scans at $\hbar\omega=$5, 6.5
and 12~meV performed in the direction transverse to ${\bf
Q}=(0.5,0.5)$ are shown. Similar to SC50, but less pronounced, we
can also see a ``resonance'' feature in SC30 in the scans of
$\hbar\omega=6.5$ and 5~meV. 

For the non-superconducting samples, there is no temperature effect observed
for data taken between 4~K and 25~K.  For NSC30, we did a constant-energy scan near $(0.5,0.5)$ only at $\hbar\omega=6.5$~meV, and the intensity is very low compared to either its
own magnetic scattering near $(0.5,0)$ or those from the other samples near
$(0.5,0.5)$.  It is clear that the low energy spin excitations
are mostly focused around $(0.5,0)$ for NSC30. The NSC45 sample has
Se doping very close to SC50, and also very similar magnetic
excitation spectrum near $(0.5,0.5)$ compared to that from the latter in its
normal state ($T=25$~K). However, with no superconducting transition, its
spectrum at low temperature ($T=4$~K) does not differ much from that at $T=25$~K.

The implications of our results are very clear for NSC30 and SC50.
For NSC30, a short-ranged static magnetic order is present at low
temperature near $(0.5,0)$, corresponding to a 3D bicollinear E-type
spin structure. Its low energy magnetic excitations are also focused
near $(0.5,0)$. With the static order present, no superconductivity
is achieved in this sample. For the SC50 sample, there is no static
order and the low energy magnetic excitation spectrum is mostly
shifted to the $(0.5,0.5)$ region, corresponding to collinear C-type
spin correlations. Similar to the situation in the
``122''~\cite{Zhaoj2009nf,Chen2009epl} or ``1111''
systems,~\cite{Clarina2008,Zhaoj2008nm,Huang2008prb,Lee2008jpsj,Zhaoj2008prb,Qiu2008prb}
this collinear configuration without static order appears to favor
superconductivity.

The results for NSC45 are more complicated. Here, with Se
doping close to SC50, the magnetic excitations near the
$(0.5,0.5)$ point are rather similar to the superconductor, except that the ``resonance''
feature is missing. Apparently, having magnetic excitations
near $(0.5,0.5)$ associated with the collinear spin configuration is
not sufficient for superconductivity to emerge. Also having a static
2D-like magnetic order with  bicollinear structure is able to
completely suppress superconductivity in this sample. 
Of course, the tetragonal crystal structure gives no energetic
distinction between the ordering wave vectors $(0.5,0)$ and
$(0.5,0.5)$, so that the magnetic configuration is relatively soft.
There is likely a mixed phase where the bicollinear and collinear
magnetic configurations coexist, on a microscopic level, similar to
that of the mixed C-E phase in manganites,~\cite{Hotta2003} as
suggested in Ref.~\onlinecite{Yinw2010}

The case for SC30 is, in fact, quite intriguing. A 2D-like
short-range static order exists at low temperature, while low energy
magnetic excitations are found both around $(0.5,0)$ and $(0.5,0.5)$
with comparable spectral weight. Therefore, the magnetic excitation
spectrum actually looks very similar to that in NSC45, yet there is
bulk superconductivity in SC30 when the static magnetic order is
also present.  Compared to NSC45, the ratio of  spectral weight near
(0.5,0.5) to that near (0.5,0) is higher in SC30, indicating a
larger volume of the sample favoring a collinear spin configuration.
The ``resonance'' occurs below $T_c$, showing an enhancement of
spectral weight only near $(0.5,0.5)$.  This indicates that
superconductivity only exists in the part of the sample with dynamic
collinear spin correlations. Although it is conceivable that the
static order and superconductivity could coexist in the same domains
as suggested by previous $\mu$SR work,~\cite{Khasanov2009} it is
also possible to have a system with macroscopic phase separation,
where the volume of local collinear or bicollinear regions are large
enough to form separate domains. In this case, the features near
$(0.5,0)$ (elastic magnetic peak and low energy magnetic
excitations), and those near $(0.5,0.5)$ come from different
regions. This scenario would be consistent with the (varying)
susceptibility results for different small pieces taken at different
locations from this sample, and agrees with results from all other
samples where static magnetic order and superconductivity do not
coexist locally.

Why would samples with similar Se content ({\it e.g.}, NSC30 vs.
SC30, NSC45 vs. SC50) show dramatically different behaviors? It is clear from previous work that the Fe content has a significant impact on the magnetism.~\cite{Baow2009prl,Wen2009} For example, higher Fe content in the
parent compound can drive the order from commensurate to
incommensurate,~\cite{Baow2009prl} while its effect has been less clear
for the superconducting region. It seems unlikely that the
excess Fe atoms act simply as isolated magnetic moments that destroy the
superconductivity, since our observations of variations in static magnetic order and low
energy spin excitations cannot be
explained in such a simple manner. 
Theoretically, it has been predicted that lowering the height of the
chalcogen (Te/Se) positions can drive the ``1:1'' system from the
bicollinear to the collinear spin
configuration,~\cite{Moon2010,Yinw2010} and Fe interstitials will
certainly have an impact on Fe-Te/Se-Fe bond lengths and bond
angles. There are several reports concerning the effect of excess Fe
on the lattice parameters~\cite{McQueen2009,
McQueen2009prb,Viennoisa2009}; our results indicate that the lattice
parameters $a$ and $c$ both decrease slightly with increased Fe
content (holding Se constant), which does not appear to be entirely
consistent with the trends reported by others.  To make further
progress on this issue, it will likely be necessary to characterize
the microstructure associated with specific
compositions.\cite{McQueen2009,Hu2009prb}

\section{Summary}

Despite uncertainties in the underlying cause of differences among our
samples, it is evident that the magnetic structure/fluctuations
are intimately correlated with the superconductivity in the 1:1 system. If the
magnetic correlations in the system favor a bicollinear (E-type) spin
configuration, which sometimes can eventually lead to static order,
superconductivity is suppressed. The collinear (C-type) spin
configuration, which is universal across the superconducting regions of
all Fe-based superconductor families, is necessary,
but not sufficient, for the
emergence of bulk superconductivity. There are cases
where magnetic correlations favoring these two configurations coexist
and compete. Overall, our results suggest that magnetic correlations are
important in the Fe-based superconductor families, and the proper tuning
of these correlations may be the key for enhancing superconductivity.

\section*{Acknowledgments}

We thank Weiguo Yin and Wei Ku for useful discussions. Work at
Brookhaven is supported by the Office of Basic Energy Sciences,
Division of Materials Science and Engineering, U.S. Department of
Energy (DOE), under Contract No.\ DE-AC02-98CH10886.  JSW and ZJX
are supported by the Center for Emergent Superconductivity, an
Energy Frontier Research Center funded by the U.S. DOE, Office of
Basic Energy Sciences. The SPINS spectrometer at the NCNR is
supported in part by the National Science Foundation under Agreement
No. DMR-0454672.


\end{document}